# COMPARISON OF DIFFERENT METHODS FOR IDENTIFICATION OF DOMINANT OSCILLATION MODE

## *KOMPARACIJA RAZLIČITIH METODA ZA IDENTIFIKACIJU DOMINANONTNIH OSCILATORNIH MODOVA*

Maja Muftić Dedović[1], Samir Avdaković[1], Adnan Mujezinović[1], Nedis Dautbašić[1]

**Abstract:** This paper introduces and compares the various techniques for identification and analysis of low frequency oscillations in a power system. Inter-area electromechanical oscillations are the focus of this paper. After multiresolution decomposition of characteristic signals, physical characteristics of system oscillations in signal components are identified and presented using the Fourier transform, Prony's method, Matrix Pencil Analysis Method, S-transform, Global Wavelet Spectrum and Hilbert Huang transform (Hilbert Marginal Spectrum) in time-frequency domain representation. The analyses were performed on real frequency signals obtained from FNET/GridEye system during the earthquake that triggered the shutdown of the North Anna Nuclear Generating Station in the east coast of the United States. In addition, according to the obtained results the proposed methods have proven to be reliable for identification of the model parameters of low-frequency oscillation in power systems. The relevant analyses are carried out in MATLAB coding environment.

**Sažetak:** U ovom su radu predstavljene i uspoređene različite tehnike prepoznavanja i analize oscilacija niskih frekvencija u elektroenergetskom sistemu. Fokus ovog rada su elektromehaničke oscilacije među zonama. Nakon multirezolucijske dekompozicije karakterističnih signala, u komponentama signala fizičke karakteristike oscilacija sistema identificiraju se i prikazuju pomoću Furijereve transformacije, Pronyjeve metode, metode Matrix Pencil, S-transformacije, globalnog valnog spektra i Hilberta Huangove transformacije (Hilbertov marginalni spektar) u vremensko frekventnoj domeni. Analize se provode nad signalima realne frekvencije, dobivenim iz FNET/GridEye sistema, tokom zemljotresa koji je pokrenuo gašenje nuklearne elektrane North Anna na istočnoj obali Sjedinjenih Država. Nadalje, pokazalo se da su predložene metode pouzdane za identificiranje niskofrekventnih dominantnih oscilacija u elektroenergetskom sistemu. Odgovarajuće analize provodene se u softeverskom paketu MATLAB.

**Ključne riječi:** oscilatorni mod, Pronijeva metoda, Matrix Pencil metoda, S-transformacija, Globalni valni spektar, Hilbertov marginalni spektar

## INTRODUCTION

The power system is a complex dynamic system exposed to constant changes and disturbances. However, disturbances can cause major problems in the power system, and the development of events and cascading shutdowns of system elements can lead to the system collapse. The stability of the power system is in the ability of system to return to its normal or stable condition after being disturbed. This paper will focus on a system exposed to major disturbance, the largest earthquake in the eastern U.S. that happened on 23 August 2011. The East Coast Earthquake caused a 1600 MW generator trip, and the measured frequencies obtained from FNET/GridEye, high dynamic accuracy Frequency Disturbance Recorders (FDRs) are used to identify the inter-area mode for this event and area [1]. The FNET is a system for monitoring grid dynamics using FDRs to collect synchronized voltage, angle, and frequency measurements. The FNET consists of 300 FDRs capable for event location and estimation of event size, wide-area frequency and angle visualization, graphic display and inter-area oscillation mode identification operated by the Power Information Technology Laboratory at the University of Tennessee [2].

Advanced WAMPC systems are expected to provide answers to many open questions regarding the stability and security of power systems. Also, the trends we recognize today in modern power systems such as integration of renewable sources, electric cars, development of telecommunications and information infrastructure, increasingly demanding regulatory requirements, new technologies in the field of metering, development of smart grids, etc., further complicate power systems. New challenges have been posed to the scientific and professional community. Finding adequate answers to challenges today requires the work of multidisciplinary teams, application of various mathematical techniques and development of new ones, and data transmission, analysis, processing and security present a particular problem and challenge. Advanced systems for monitoring, protection and management of power systems are expected to be able (among other things) to efficiently identify system disturbances, determine the location of the disturbance, identify low-frequency electromechanical oscillations and their character, assess the active power imbalance in the system, be able to take automated actions to prevent cascading propagation of disturbances and / or minimise the impact of disturbances on the system, prevent unnecessary protection actions, minimise the possibility of human error, etc [3].

[1]Faculty of Electrical Engineering, BIH
maja.muftic-dedovic@etf.unsa.ba

Low-frequency electromechanical oscillations in power systems are a common phenomenon and can be caused by various operating events. Most oscillations are damped in nature, but undamped oscillations can lead a system to collapse. Therefore, the identification and monitoring of low frequency oscillations in power systems is a very important aspect of modern systems for monitoring, protection and control. Low-frequency electromechanical modes that occur in the power system can be classified with the impact they have on the system component as well as on the effects they produce. The basic division according to [4] are global mode (low-frequency mode of 0,05-0,2 Hz), local area mode (frequency range is 1 – 2 Hz), intra-plant mode (the range of modal frequency is 1,5-2,5 Hz), inter-area mode (0,25-1 Hz), torsional modes between rotating plant (10-46 Hz) and control mode oscillations. In this paper inter-area modes are identified according to various techniques, and the results are compared.

This paper is related to the phenomena of dynamic behaviour of power systems in the frequency range up to 5 Hz. This frequency range represents the range of low-frequency electromechanical oscillations. Different approaches were presented in literature to identify low-frequency electromechanical oscillations. Prony's method is widely used in literature for identification of low frequency oscillations in power system and can be found in [5], [6]. Because Prony's method's results are strongly influenced by noise, implementation of Fast-Fourier transform (FFT) for LFO characterization can be found in [7], [8]. In [9], [10], Matrix Pencil Analysis Method for identification of oscillation modes is used. Ref. [11] presented S-transform for estimating the low frequency modes in a given ring down signal. The Wavelet transform analysis of disturbance identification and the analysis and identification of low-frequency electromechanical oscillations (LFEOs) in power systems is presented in [12]. A method for LFO identification based on Hilbert Huang transform was proposed in [13], [14].

This paper introduces the techniques for identification and analysis of low frequency oscillations in power system. After multiresolution decomposition of characteristic signals, physical characteristics of system oscillations in signal components are identified and presented using the Fourier transform, Prony's method, Matrix Pencil Analysis Method, S-transform, Global Wavelet Spectrum and Hilbert Huang transform (Hilbert Marginal Spectrum) in time-frequency domain representation. The relevant analyses are carried out in MATLAB coding environment.

The remainder of this paper is organized as follows: Description of the disturbance in August 2011 is presented in Section 1. In Section 2, basic theory of applied approaches for identification of inter-area modes are presented. Practical application results and analysis of identification of low frequency oscillations on the actual frequency information obtained from WAMS established as a part of FNET/GridEye system are given in Sections 3, while Section 4 offers the conclusion.

## 1. DESCRIPTION OF THE DISTURBANCE IN AUGUST 2011

First of all, the measured frequency signals from FNET/GridEye after disturbance in North America are presented in Figure 1. There are approximately 80 FDRs deployed throughout North America. The analyses are performed on real frequency signals obtained from FNET/GridEye system during the earthquake that triggered the shutdown of the North Anna Nuclear Generating Station resulting in 1600 MW of generation loss in the Eastern Interconnect [15]. Frequencies obtained from FDRs are used for identification of inter-area modes and to compare obtained results according to FNET inter-area oscillation modal analysis results presented in [1]. In [1] is concluded that 0.2 Hz is the major mode shared by all the monitoring locations and this value is used as measure for validation of results accuracy in accordance with a particular technique used for identification of dominant modes. For further analyses are taken frequency signals from the north-western (Winnipeg), the south-eastern region (Pansacola) and the north-eastern region (Boston).

Figure 1. shows frequency oscillations during the largest earthquake in the eastern U.S. for 48 FDR units.

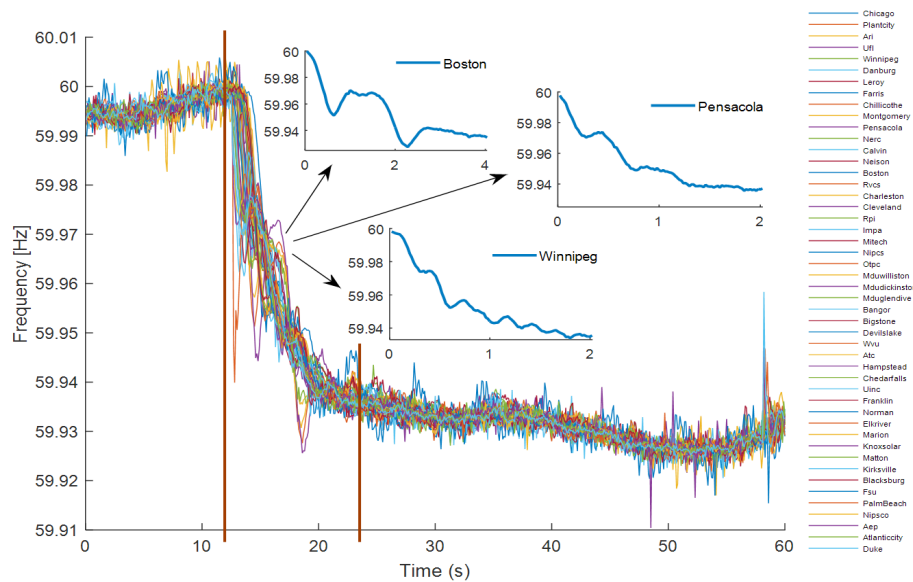

Figure 1: Frequency oscillations recorded by FDRs during the largest earthquake in the eastern U.S.

## 2. APPLIED METHODS FOR OSCILLATION MODES IDENTIFICATION

In this section various techniques used for identification of low frequency oscillations in power system are presented and briefly explained.

### 2.1. Fourier Transform

The Fourier transform provides a good insight into the frequency content of the signal. Also, the Fourier transform is one of the most commonly used signal processing tools. It converts the signal (function in the time domain) into a record in the frequency domain, thus obtaining information about the recording of a given signal using a sinusoid, i.e. which frequencies appear and at what intensity, but information about time is lost, i.e. when which frequency occurred. When analysing signals that are stationary or periodic, the Fourier transform is sufficient for analysis. However, very often non-stationary and non-periodic signals appear in the power system and information on time and frequency is needed, i.e. it is necessary to have information when certain frequencies occurred. Accordingly, it is necessary to implement time-frequency analysis. According to Heisenberg's uncertainty principle, not all frequencies and time components can be known at the same time, but the range of frequencies for a given moment in time can. One approach to time-frequency analysis is short-time Fourier transform (STFT). The mathematical description and steps of FFT can be found in detail in the following references [16]-[19]. In this section by applying FFT function in the MATLAB over all frequencies from 48 FDR units the goal is to yield the dominant low frequency modes of the oscillations during the fault conditions. The parameters of the fast Fourier Transform Gaussian Window Factor are chosen to be 0.3.The following figure shows the results of processing the input measurements shown in Figure1. using the described method FT. Also Figure 2 emphasis FDR units from different regions of EI that will be analysed by other applied methods for oscillation modes identification.

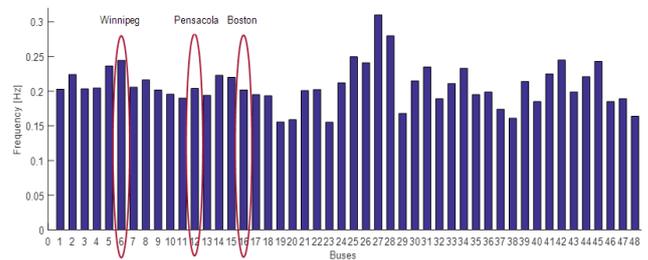

Figure 2: Dominant frequencies of 48 oscillation for the event on August 23, 2011 obtained by FT method

### 2.2. Prony's Method

Prony's method is a signal processing technique that extends Fourier analysis and also belongs to the group of parametric methods. Prony's method is developed by Gaspard Riche de Prony, in 1795 [20]. By this method the equally sampled data could be approximated by exponential functions and behave like least squares linear prediction algorithm which is applied for AR (Autoregressive) and ARMA (Autoregressive moving average) parameter detection [17]. Actually, the signal is modeled as a sum of damped, complex exponentials or equivalently decaying sinusoidals. The main objective is the precise estimation of signal components parameters: frequency, amplitude and phase, as well as coefficients suppression. The method consists in presenting the signal as a linear combination of exponential functions, which for real signals can be expressed by the equation (1).

$$\widehat{x}_n = \sum_{k=1}^{p/2} 2a_k e^{\left[\alpha_k(n-1)T\right]} \cos\left[2\pi f_k (n-1)T + \theta_k\right] \quad (1)$$

where:
$n=1, 2, …, N$,
$N$ - signal length (number of samples),
$p$ - number of exponential components (model order),
$T$ - sampling period (s),
$a_k$ - amplitude of the k-th,
$\alpha_k$ - damping factor (s$^{-1}$),
$f_k$ - sinusoidal frequency (Hz),
$\theta_k$ - initial phase of the sinusoidal component (rad).

Further, explanation of Prony analysis and implementation in identification of dominant oscillator modes can be found in [21], [22].

For the purposes of this paper Prony energy spectrum is applied on the measured frequency from three FDR units in North America during disturbance. Results are plotted on Figure 3.

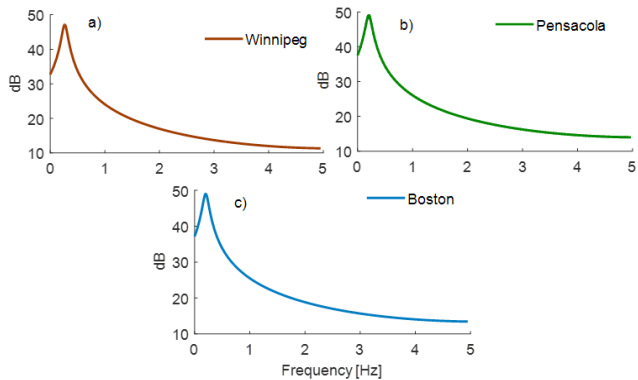

Figure 3: Prony energy spectrum a) Winnipeg b) Pensacola c) Boston

### 2.3. Matrix Pencils Method

The Matrix Pencil method (MPM) is used to decompose the signal into a preselected number of exponentially decaying sinusoids [23]. The details of the method algorithm are given in [24], [25]. Matrix pencil method with Hankel total least squares method is used in estimation of the electromechanical modes using measured frequency data. Also, MPM is computationally very efficient and less sensitive to signal with noise. This method is implemented over the data window and for each window, resulting mode energy is generated and the mode with the highest energy is selected to be dominant for the corresponding data window. This procedure is repeated by moving the data window along the entire length of the signal, generating energy of different values and many possible modal frequencies. Once all these steps have been performed for modal decomposition the fit and reconstruction of the signal is ready to be executed.

In Table I. presents the estimated values of the dominant oscillator mode accurately at 0.257 Hz, 0.208 Hz and 0.201 Hz, respectively for Winnipeg, Pensacola and Boston FDR units. Figure 4 shows comparison of the signal estimated by the Matrix Pencil method with the original measured signals according to selected FDR unit.

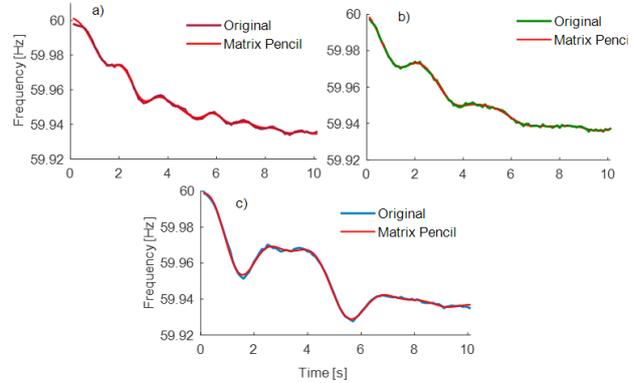

Figure 4: Signal reconstruction by Matrix Pencil method a) Winnipeg b) Pensacola c) Boston

### 2.4. S-transform

The S-transform is an invertible time-frequency analysis technique that combines elements of wavelet transforms and short-time Fourier transforms, but with a Gaussian-shaped window whose width scales inversely with frequency [26]. The expression of the S-transform is [27]:

$$S(\tau - t) = \int_{-\infty}^{\infty} S(t) \left\{ \frac{|f|}{\sqrt{2\pi}} e^{\left[\frac{-f^2(\tau-t)^2}{2}\right]} e^{-2\pi i f t} \right\} dt \quad (2)$$

where:
$S(t)$ - signal,
$\tau$ - parameter which controls the position of the Gaussian window.

The S-transform is the time-frequency representation that is similar to wavelet and STFT. It combines a frequency dependent resolution of the time-frequency space with the precise referenced local phase information [28]. The S-transform uses a Gaussian-shaped window to localize the complex Fourier sinusoid; its width scales with frequency, unlike the STFT and analogously to the wavelet transform. The time-frequency resolution is improved with S-transform which is also easy to manipulate for interpretations. Figure 5. shows the time-frequency representation of the dominant oscillation mode using the S-transform.

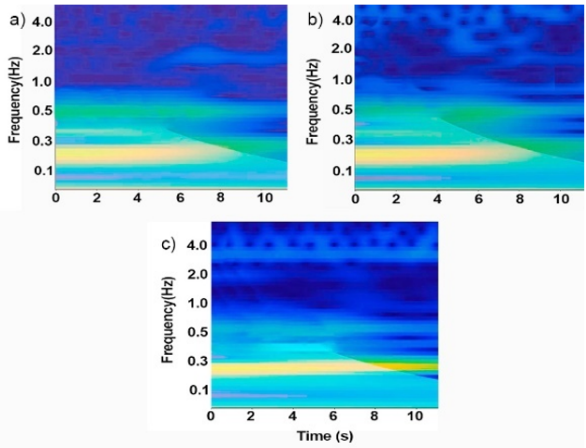

Figure 5: The time-frequency representation using the S-transform a) Winnipeg b) Pensacola c) Boston

## 2.5. Global Wavelet Spectrum

The Continuous Wavelet transformation (CWT) of discrete signals is defined as the convolution of time series or signals by scaled and translated versions of the conjugate-complex wavelet function, where the wavelet function is a frequency band pass filter on the time series or signal ($x_n$, n=0,1,2,…, N-1). As $W_n^x$ represents the wavelet transformation of the x signal, in energy terms it is easy to define the wavelet power spectrum (WPS) of the signal as $|W_n^x|^2$ represents the local variance of x, while the complex argument of WPS represents the local phase. If necessary, it is easy to find the global wavelet transform power spectrum (GWS), which represents the time average of the wavelet transform power spectrum and is defined as $\int_{-\infty}^{\infty} |W_n^x|^2 dt$. WPS provides a large amount of useful information, while GWS can in practical terms provide useful comparisons of spectral power on different scales. More details about Wavelet transform and Global Power spectrum can be found in [29], [30]. Figure 6. presents GWS of all measured power system frequencies during large disturbance.

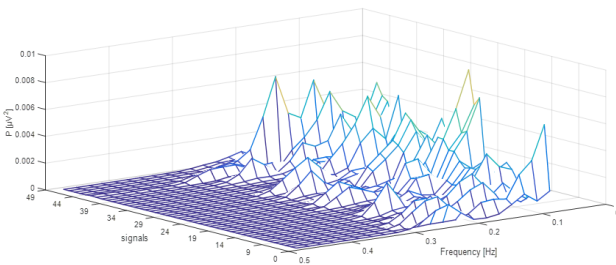

Figure 6: GWS results of a signals from Figure 1

## 2.6. Hilbert Marginal Spectrum

The Hilbert Huang transform consists of two parts: Empirical Mode Decomposition (EMD) and Hilbert spectral analysis. Decomposition is based on the characteristics of the scale and therefore it can be used for the analysis of nonlinear and nonstationary signals, and for three-dimensional signal presentation in the domains of time, frequency and energy. Although this method is completely empirical, the results it can give can be quite powerful.

The fundamental and first part of HHT is the Empirical Mode Decomposition of the signal. This procedure firstly decomposes the signal into simple intrinsic mode functions (IMF). Its peculiarity lies in the fact that it can decompose any signals into a finite and small number of components that are the parts of the IMF. Each IMF function is a simple harmonic function whose amplitude and frequency can also be functions of time. The number of extremes and zero-points of the function along the entire length of the signal can differ by a maximum of one differ by a maximum of one. The requirement of the IMF is that the mean value of the envelope, which is defined by local maxima and local minima, is equal to zero at any point of time. The signal first identifies all local extremes - maxima and minima. The local maxima are then linked by a 'cubic spline' function thus obtaining an upper envelope. By connecting all local minima with the same function, the lower envelope is obtained. After satisfying the properties over each of these IMF functions, the Hilbert transformation is performed [31], [32].

If $H[c_i(t)]$ denotes HHT of $i$-th IMF component, then the analytical form of the signal $c_i(t)$ is formed by:

$$z_i(t) = c_i(t) + jH[c_i(t)] = a_i(t)e^{j\varphi_i(t)} \quad (3)$$

$$a_i(t) = \sqrt{c_i(t)^2 + H[c_i(t)]^2} \quad (4)$$

$$\phi_i(t) = \arctan \frac{H[c_i(t)]}{c_i(t)} \quad (5)$$

The instantaneous frequency is as shown below:

$$w_i(t) = \frac{d\varphi_i(t)}{dt} \quad (6)$$

If $a_i(t)$, $\theta_i(t)$ and $\omega_i(t)$ denote amplitude, phase and instantaneous frequency of signal $z_i(t)$ respectively, original signal is as shown below:

$$x(t) = H[\omega,t] = \text{Re}\left\{ a_i(t)\exp(j \cdot \int \omega_i(t)dt) \right\} \quad (7)$$

The instantaneous frequency is as shown below:

$$h(\omega) = \int_0^T H[\omega,t]dt \quad (8)$$

where, $T$ is duration of signal [14].

In this paper, HMS obtained by EMD is applied for the identification of the dominant oscillation mode in a power system. The results of the HMS approach are shown in Figure 7, where can also be concluded that the HMS approach for signal from Figure 1. clearly identifies the dominant oscillatory mode of 0.2 Hz. Figure 7. shows plotted power spectra for all and for selected FDR units Winnipeg, Pensacola, Boston as illustrative examples in this paper.

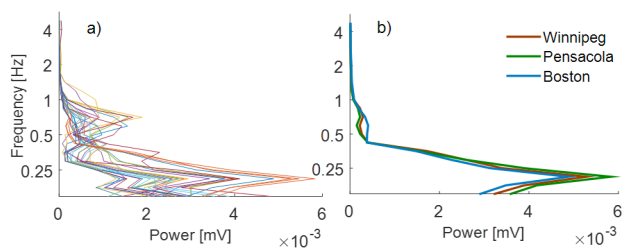

Figure 7: Power spectra for
a) all FDR units b) FDR units: Winnipeg, Pensacola, Boston

## 3. RESULTS

In this paper the MATLAB codes for FT, Prony's method, MPM, S-transform, GWS and HMS are utilized to extract the power system dominant low-frequency eletromechanical oscillations. In the previous section, the figures were selected as indicators of the applied methods accuracy, and in this section the numerical values of the dominant oscillatory modes identification results are singled out. The results are presented in Table I.

Table I: Dominant oscillatory modes

| Method | FDR unit | | |
|---|---|---|---|
| | Winnipeg | Pensacola | Boston |
| Fourier transform | 0.2445 | 0.204 | 0.2018 |
| Prony's method | 0.2661 | 0.203 | 0.2004 |
| Matrix Pencil Method | 0.2355 | 0.1987 | 0.199 |
| S-transform | 0.252 | 0.2102 | 0.2311 |
| Global Wavelet Spectrum | 0.2367 | 0.2184 | 0.2 |
| Hilbert Marginal Spectrum | 0.2125 | 0.213 | 0.213 |

The signal processing based techniques discussed in this paper provides satisfactory performance in detecting the low frequency modes. The results from Table I obtained by all applied methods are compared to inter-area oscillation modal analysis results presented in [1]. According to reference [1] 0.2 Hz is the dominant oscillation mode shared by all the monitoring locations during large disturbance in North America. Also, the results from Table I clearly show great accuracy according to all applied methods for dominant mode identification. The frequencies obtained from these methods are more or less uniform and consistent, implying the efficacy and accuracy of the signal processing methods used in this paper.

## 4. CONCLUSION

The efficiency of the proposed methods are demonstrated using measured signal, a practical case study to estimate the dominant low frequency oscillatory modes. The East Coast U.S. Earthquake that happened on August 23, 2011 caused a 1600 MW generator trip, and the measured frequencies obtained from FNET/GridEye are used for analyses and comparison of different methods for the extraction of power system dominant low-frequency eletromechanical oscillations. The six signal processing approaches, Fourier transform, Prony's method, Matrix Pencil Analysis Method, S-transform, Global Wavelet Spectrum and Hilbert Marginal Spectrum clearly distinguish dominant oscillation modes and can provide very useful information for the operators of power systems. It is important to emphasise that multi oscillation modes are excited after analyses for some FDR units, but only mode with the highest energy is extinguished as a major oscillation mode. Also, the computation time required by the proposed methods is low in relation to the dynamics of low-frequency oscillations. With a results from the numerous methods for identifying and analyzing oscillating events, the models dynamic characteristics of all major networks can be well understood. Thus, more accurate results lead to more successful planning and management and can be implemented using FNET oscillation data.

## BIOGRAPHY

**Maja Muftić Dedović** was born in Sarajevo, Bosnia and Herzegovina. She received the B.Eng. and M.Sc. degrees in electrical engineering from the Faculty of Electrical Engineering, University of Sarajevo. Currently she is pursuing the Ph.D. degree at the Faculty of Electrical Engineering, University of Sarajevo. She works at the Faculty of Electrical Engineering, University of Sarajevo as a teaching assistant. Her research interests are power system analysis, power system dynamics and stability, WAMPCS and signal processing.

**Samir Avdaković** was born in Doboj, Bosnia and Herzegovina. He received the Ph.D. degree in electrical engineering from the Faculty of Electrical Engineering, University of Tuzla in 2012. He works at the Department for Strategic Development in EPC Elektroprivreda B&H. His research interests are power system analysis, power system dynamics and stability, WAMPCS and signal processing.

**Adnan Mujezinović** received his M.Sc. and Ph.D. degrees in electrical engineering from University of Sarajevo (Bosnia and Herzegovina), in 2011 and 2017, respectively. Currently, he is an assistant Professor at the Faculty of Electrical Engineering, University of Sarajevo. His topics of interest are numerical calculations of electromagnetic fields, cathodic protection and grounding systems.

**Nedis Dautbasić** was born in Srebrenica, Bosnia and Herzegovina. He received the B.Eng. and M.Sc. degrees in electrical engineering from the Faculty of Electrical Engineering, University of Sarajevo. Currently he is pursuing the Ph.D. degree at the International Burch University. He works at the Faculty of Electrical Engineering, University of Sarajevo as a teaching assistant. His research interests are power system analysis, power system dynamics and stability, WAMPCS and signal processing.